\documentclass[10pt]{article}
\usepackage[cp866]{inputenc}
\begin{document}

\centerline {{\large\bf Analysis of the equations of mathematical physics}}
\centerline {{\large\bf and foundations of field theories with the help}}
\centerline {{\large\bf of skew-symmetric differential 
forms}}

\centerline {\bf L.I. Petrova}
\centerline{{\it Moscow State University, Russia, e-mail: ptr@cs.msu.su}}
\renewcommand{\abstractname}{Abstract}
\begin{abstract}

In the paper it is shown that, even without a knowledge of the concrete 
form of the equations of mathematical physics and field theories, with 
the help of skew-symmetric differential forms one can see specific features 
of the equations of mathematical physics, the relation between mathematical 
physics and field theory, to understand the mechanism of evolutionary 
processes that develop in material media and lead to emergency of 
physical structures forming physical fields. This discloses a physical 
meaning of such concepts like "conservation laws", "postulates" and 
"causality" and gives answers to many principal questions of mathematical 
physics and general field theory.

In present paper, beside the exterior forms, the skew-symmetric 
differential forms, whose basis (in contrast to the exterior forms) 
are deforming manifolds, are used.
Mathematical apparatus of such differential forms
(which were named evolutionary ones) 
includes nontraditional elements like nonidentical relations and 
degenerate transformations and this enables one to describe discrete 
transitions, quantum steps, evolutionary processes, and generation of 
various structures.

\end{abstract}

{\large\bf Introduction}

Skew-symmetric differential forms possess unique properties that enable 
one to carry out a qualitative investigation of the equations of 
mathematical physics and the foundations  of field theories. They can 
describe a conjugacy of various operators and objects 
(derivatives, differential equations, and so on).

Such a potentiality of skew-symmetric differential forms
is due to the fact that skew-symmetric differential forms,
as opposed to differential equations, deal with differentials
and differential expressions rather than with derivatives.

In the paper, beside the exterior skew-symmetric differential
forms that can describe conjugated objects, the skew-symmetric differential
forms, whose basis (in contrast
to the exterior forms) are deforming manifolds, are used. Such 
skew-symmetric differential forms, which were named evolutionary ones, 
can describe the process of conjugating objects and obtaining conjugated 
objects - the closed exterior forms. Evolutionary forms are obtained 
from the equations modelling physical processes, and therefore, they 
possess evolutionary properties.

The physical meaning of exterior skew-symmetric differential
forms is connected with the fact that they correspond to conservation 
laws. Closed (inexact) exterior forms and relevant dual forms compose 
conjugated objects (a differential-geometrical structure). Just such 
conjugated objects, which are invariant ones, correspond to conservation 
laws. These are conservation laws for physical fields. The physical
structures that form physical fields are just such conjugated objects.

The theory of closed exterior forms lies at the basis of field theories
(the theories describing physical fields).
The invariant properties of exterior forms explicitly or implicitly
manifest themselves essentially in all formalisms of field theory, such
as the Hamilton formalism, tensor approaches, group methods, quantum
mechanics equations, the Yang-Mills theory and others.
The gauge transformations (unitary, gradient and so on), the gauge symmetries
and the identical relations of field theories are transformations, symmetries
and relations of the theory of closed exterior forms.

In the paper it will be shown that the closed exterior forms, whose properties
lie at the basis of field theories, are obtained from evolutionary forms related to
the equations of mathematical physics. This discloses a relation between
mathematical physics and the invariant field theories.

Evolutionary forms, as well as closed exterior forms, reflect the properties
of conservation laws. However, they are conservation laws for material media.
They are balance (differential) conservation laws. They are conservation laws
for energy, linear momentum, angular momentum, and mass.

Below, on the basis of the properties of evolutionary forms it will be shown
the noncommutativity of the balance conservation laws and their controlling
role in the evolutionary processes developing in material media. In will be
shown that such processes lead to origination of physical structures from which
physical fields are formatted. During this the origination of physical structures
in these processes reveals as an emergency of such formations like waves, vortices, turbulent
pulsations, massless particles, and so on.

Such results firstly proves that material media generate
physical fields. And secondly, they explain the nature of turbulence and various
types of instabilities developed in material media.

Below it will be shown that the parameters of exterior and evolutionary forms
allow to introduce a classification of physical fields and interactions.

The methods of investigating concrete material systems and physical fields
on the basis of the exterior and evolutionary differential forms
are demonstrated by the examples that are presented in Appendices 2.

\bigskip 

The existence of evolutionary skew-symmetric differential forms has
been established by the author while studying the problems of stability.
Mathematical apparatus of evolutionary skew-symmetric differential forms 
includes nontraditional elements like nonidentical relations and 
degenerate transformations and this enables one to describe discrete 
transitions, quantum steps, evolutionary processes, and generation of 
various structures. Such mathematical apparatus is beyond the scope of 
existing physical and mathematical formalisms, and this
makes difficulties in presentation and perception of this matter.

\bigskip
{\large\bf Closed exterior forms. Conservation laws}

It is known that the exterior differential form of degree $p$ ($p$-form)
can be written as [1,2]
$$
\theta^p=\sum_{i_1\dots i_p}a_{i_1\dots i_p}dx^{i_1}\wedge
dx^{i_2}\wedge\dots \wedge dx^n\quad 0\leq p\leq n\eqno(1)
$$
Here $a_{i_1\dots i_p}$ are functions of the variables $x^1$,
$x^2$, \dots, $x^n$, $n$ is the dimension of space,
$\wedge$ is the operator of exterior multiplication, $dx^i$,
$dx^{i}\wedge dx^{j}$, $dx^{i}\wedge dx^{j}\wedge dx^{k}$, \dots\
is the local basis which satisfies the condition of exterior
multiplication (the condition of skew-symmetry). 

The differential of the exterior form $\theta^p$ is expressed as
$$
d\theta^p=\sum_{i_1\dots i_p}da_{i_1\dots
i_p}\wedge dx^{i_1}\wedge dx^{i_2}\wedge \dots \wedge dx^{i_p} \eqno(2)
$$

The physical sense have the closed exterior forms. [Below we present only
the data on exterior forms which are necessary for further
presentation].

From the closure condition of the exterior form $\theta^p$:
$$
d\theta^p=0\eqno(3)
$$
one can see that the closed exterior form $\theta^p$ is a conserved
quantity. This means that this can correspond to a conservation law,
namely, to some conservative quantity.

If the form is closed only on pseudostructure, i.e. this form is
a closed inexact one, the closure conditions are written as
$$
d_\pi\theta^p=0\eqno(4)
$$
$$
d_\pi{}^*\theta^p=0\eqno(5)
$$
where ${}^*\theta^p$ is the dual form.

Condition (5), i.e. the closure condition for dual form, specifies
the pseudostructure $\pi$.
\{Cohomology (de Rham cohomology, singular cohomology 
and so on), sections of cotangent bundles, the surfaces
eikonals, potential surfaces, pseudo-Riemannian and pseudo-Euclidean
spaces, and others are examples of the psedustructures
and manifolds that are formed by pseudostructures.\}

From conditions (4) and (5) one can see the following. The dual form
(pseudostructure) and closed inexact form (conservative quantity)
made up a conjugated conservative object that can also correspond to
some conservation law.
The conservative object, which corresponds to the conservation law,
is a differential-geometrical structure. (Such
differential-geometrical structures are examples of G-structures.)
The physical structures, which made up physical
fields, and corresponding conservation laws are just such structures.

\bigskip
{\bf Properties of closed exterior differential forms.}

1. {\it Invariance.}

It is known that the closed exact form is
a differential of the form of lower degree:
$$
\theta^p=d\theta^{p-1}\eqno(6)
$$
Closed inexact form is also a differential, and yet not a total one but
an interior on pseudostructure
$$
\theta^p_\pi=d_\pi\theta^{p-1}\eqno(7)
$$

Since the closed form is a differential (a total one if the form is exact,
or an interior one on the pseudostructure if the form is inexact), it is
obvious that the closed form turns out to be invariant under all
transformations that conserve the differential. The unitary transformations
($0$-form), the tangent and canonical transformations ($1$-form),
the gradient transformations ($2$-form) and so on are examples of
such transformations.
{\it These are gauge transformations for spinor, scalar, vector, tensor
($3$-form) fields}. It should be pointed out
that just such transformations are used in field theory.

With the invariance of closed forms it is connected the covariance of relevant
dual forms.

2. {\it Conjugacy. Duality. Symmetries.}

The closure of exterior differential forms is the result of conjugating the
elements of exterior or dual forms. The closure property of the exterior
form means that any objects, namely, the elements of exterior form, the
components of elements, the elements of the form differential, the exterior and
dual forms, the forms of sequential degrees  and others, turn out to be
conjugated.

With the conjugacy it is connected the duality.

The example of a duality having physical sense: the closed exterior form
is a conservative quantity corresponding to conservation law, and the
closed form (as the differential) can correspond to
a certain potential force. (Below it will be shown in
respect to what the closed form manifests itself as a potential
force and with what the conservative physical quantity is connected).

The conjugacy is possible if there is one or another type of symmetry.

The gauge symmetries, which are interior symmetries of field theory and
with which gauge transformations are connected, are symmetries
of closed exterior differential forms. The conservation laws for 
physical fields are connected with such interior symmetries.

\bigskip
{\bf Identical relations of exterior forms.}

Since the conjugacy is a certain connection between two operators or
mathematical objects, it is evident that, to express the conjugacy
mathematically, it can be used relations.
These are identical relations.

The identical relations express the fact that each closed exterior
form is the differential of some exterior form (with the degree less
by one). In general form such an identical relation can be written as
$$
d _{\pi}\varphi=\theta _{\pi}^p\eqno(8)
$$

In this relation the form in the right-hand side has to be a
{\it closed} one.

Identical relations of exterior differential forms are a mathematical
expression of various types of conjugacy that leads to closed exterior
forms.

Such relations like the Poincare invariant, vector and tensor identical
relations, the Cauchi-Riemann conditions, canonical relations, the
integral relations by Stokes or Gauss-Ostrogradskii, the thermodynamic
relations, the eikonal relations, and so on are examples of identical
relations of closed exterior forms that have either the form of relation 
(8) or its differential or integral analogs.

One can see that identical relations of closed exterior differential
forms make itself evident in various branches of physics and mathematics.

Below the mathematical and physical meaning of these relations will
be disclosed with the help of evolutionary forms.

\bigskip
{\large\bf The analysis of field theories with the help of closed exterior forms.}

The properties of closed exterior differential forms correspond to the
conservation laws for physical fields. Therefore, the mathematical 
principles of the theory of closed exterior differential forms lie at
the basis of existing field theories.
\{The physical fields [3] are a special form of the substance, they are 
carriers of various interactions such as electromagnetic, gravitational, 
wave, nuclear and other kinds ofinteractions. The conservation laws for 
physical fields are those that state an existence of conservative 
physical quantities or objects.\}

The properties of closed exterior and dual forms, namely, invariance,
covariance, conjugacy, and duality, lie at the basis of the group and 
structural properties of field theory.

The nondegenerate transformations of field theory are transformations of
closed exterior forms. As it has been pointed out, the gauge
transformations like the unitary, tangent, canonical, gradient and other
gauge transformations are such transformations. These are transformations
conserving the differential. Applications of nondegenerate
transformations to identical relations enables one to find new closed
exterior forms and, hence, to find new physical structures.

The gauge, i.e. internal, symmetries of the field theory equations are
those of closed exterior forms.

The nondegenerate transformations of exterior differential forms lie at
the basis of field theory operators.
If, in addition to the exterior differential, we introduce
the following operators: (1) $\delta$ for transformations that convert
the form of $(p+1)$ degree into the form of $p$ degree, (2) $\delta'$
for cotangent transformations, (3) $\Delta$ for the
$d\delta-\delta d$ transformation, (4) $\Delta'$ for the $d\delta'-\delta'd$
transformation, one can write down the operators in the field
theory equations in terms of these operators that act on the exterior
differential forms. The operator $\delta$ corresponds to Green's
operator, $\delta'$ does to the canonical transformation operator,
$\Delta$ does to the d'Alembert operator in 4-dimensional space, and
$\Delta'$ corresponds to the Laplace operator. 

It can be shown that the equations of existing field theories are
those obtained on the basis of the properties of the exterior form
theory.
The Hamilton formalism is based on the properties of closed exterior 
form of the first degree and corresponding dual form.  
The closed exterior differential form $ds=-Hdt+p_j dq_j$
(the Poincare invariant) corresponds to the field equation related to
the Hamilton system.
The Schr\H{o}dinger equation in quantum mechanics is an analog
to field equation, where the conjugated coordinates are changed by
operators. It is evident that the closed exterior form of zero degree 
(and dual form) correspond to quantum mechanics. Dirac's {\it bra-} and 
{\it cket}- vectors constitute a closed exterior form of zero degree [4].
The properties of closed exterior form of the second degree (and dual 
form) lie at the basis of the electromagnetic field equations. 
The Maxwell equations may be written as [5]
$d\theta^2=0$, $d^*\theta^2=0$, where $\theta^2=
\frac{1}{2}F_{\mu\nu}dx^\mu dx^\nu$ (here $F_{\mu\nu}$ is the strength 
tensor).
Closed exterior and dual forms of the third degree correspond to the
gravitational field. (However, to the physical field of given type it can
be assigned closed forms of less degree. In particular, to the Einstein
equation for gravitational field it is assigned the first degree closed
form, although it was pointed out that the type of a field with
the third degree closed form corresponds to the gravitational field.)

The connection between the field theory equations and gauge transformations
used in field theories with closed exterior forms of appropriate degrees
shows that there exists a commonness between field theories
describing physical fields of different types. This can serve as an approach
to constructing the unified field theory. This connection shows that it is possible
to introduce a classification of physical fields according to the degree
of closed exterior form. Such a classification also exists for interactions
(see below).

(But within the framework of only exterior differential forms one cannot
understand how this classification is explained. This can be elucidated
only by application of evolutionary differential forms.)

And here it should underline that the field theories are based on the
properties of closed {\it inexact} forms. This is explained by the fact
that only inexact exterior forms can correspond to the physical
structures that form physical fields.  The condition that the closed
exterior forms,
which constitute the basis of field theory equations, are inexact ones
reveals in the fact that essentially all existing field theories include
a certain elements of noninvariance. Such elements of noninvariance are,
for example, nonzero value of the curvature tensor in Einstein's
theory [6], the indeterminacy principle in Heisenberg's theory, the
torsion in the theory by Weyl [6], the Lorentz force in electromagnetic
theory [7], an absence of general integrability of the Schr\H{o}dinger
equations, the Lagrange function in the variational methods, an absence
of the identical integrability of the mathematical physics equations and
that of identical covariance of the tensor equations,
and so on. Only if we assume the elements of noncovariance, we can obtain
closed {\it inexact} forms that correspond to physical structures.

And yet, the existing field theories are invariant ones because they are
provided with additional conditions under which the invariance or
covariance requirements have to be satisfied. It is possible to show
that these conditions are the closure conditions of exterior or dual
forms. Examples of such conditions may be the identity relations:
canonical relations in the Schr\H{o}dinger equations, gauge invariance
in electromagnetic theory, commutator relations in the Heisenberg theory,
symmetric connectednesses, identity relations by Bianchi in the Einstein
theory, cotangent bundles in the Yang-Mills theory, the Hamilton
function in the variational methods, the covariance conditions in the
tensor methods, etc.

It is known that the equations of existing field theories and the
mathematical formalisms of field theories have been
obtained on the basis of postulates. One can see that these postulates
are obtained from the closure conditions of inexact exterior forms.

\bigskip
Thus one can see that the properties and mathematical apparatus of
closed exterior forms made up the basis of existing field theories.

And here it arises the question of  how closed inexact exterior forms,
which correspond to physical structures and reflect the properties of
conservation laws and on whose properties field theories are based, are
obtained. This gives the answers to the following
questions: (a) how the physical structures, from which physical fields are
formatted, originate; (b) what generates physical structures; (c) how the
process of generation proceeds, and (d) what is responsible for such processes?
That is, this enables one to understand the heart of physical evolutionary
processes and their causality. This has to explain both the internal connection
between different physical fields and their classification.

Below it will be shown that the closed inexact exterior forms can be
obtained from the evolutionary forms.

\bigskip
{\large\bf Distinction between exterior and evolutionary forms}

Skew-symmetric differential forms, that the author named evolutionary ones,
differ in their properties from exterior forms.
The distinction between exterior and evolutionary skew-symmetric
differential forms is connected with the properties of manifolds
on which skew-symmetric forms are defined.

It is known that the exterior differential forms are skew-symmetric
differential forms whose basis are differentiable manifolds or they can
be manifolds with structures of any type [2,8].
(Such manifolds have one common property, namely, they locally admit
one-to-one mapping into the Euclidean subspaces and into other manifolds
or submanifolds of the same dimension [8]).

While describing the evolutionary processes in material systems
(material media) one is forced to deal with manifolds which do not allow one-to-one
mapping described above. Lagrangian manifolds and tangent manifolds of
differential equations describing physical processes can be examples of
deforming manifolds.

Such manifolds are those constructed of
trajectories of the material system elements (particles). These manifolds,
which can be called accompanying manifolds, are deforming variable
manifolds.

The skew-symmetric differential forms defined on these manifolds are
evolutionary ones. The coefficients of these differential forms and the
characteristics of corresponding manifolds are interconnected and are
varied as functions of evolutionary variables.

A distinction between manifolds on which exterior and evolutionary forms are
defined relates to the properties of metric forms of these manifolds.

Below we present some information about the manifolds on which skew-symmetrical
differential forms are defined.

\bigskip
{\bf Some properties of manifolds.}

Assume that on the manifold one can set the
coordinate system with base vectors $\mathbf{e}_\mu$ and define
the metric forms of manifold [6]: $(\mathbf{e}_\mu\mathbf{e}_\nu)$,
$(\mathbf{e}_\mu dx^\mu)$, $(d\mathbf{e}_\mu)$. The metric forms
and their commutators define the metric and differential
characteristics of the manifold.

If metric forms are closed
(the commutators are equal to zero), the metric is defined
$g_{\mu\nu}=(\mathbf{e}_\mu\mathbf{e}_\nu)$, and the results of
translation over manifold of the point
$d\mathbf{M}=(\mathbf{e}_\mu dx^\mu)$ and of the unit frame
$d\mathbf{A}=(d\mathbf{e}_\mu)$ prove to be independent of the
curve shape (the path of integration).

To describe the manifold differential characteristics
and, correspondingly, the metric form commutators, one can use
connectednesses [2,6].
If the components of metric form can be
expressed in terms of connectedness $\Gamma^\rho_{\mu\nu}$ [6],
the expressions $\Gamma^\rho_{\mu\nu}$,
$(\Gamma^\rho_{\mu\nu}-\Gamma^\rho_{\nu\mu})$ and
$R^\mu_{\nu\rho\sigma}$ are components of the commutators of the metric
forms with zeroth-, first-, and third degrees. (The commutator of
the second degree metric form is written down in a more complex
manner [6], and therefore it is not presented here).

The closed metric forms define the manifold structure.
And the commutators of metric forms
define the manifold differential characteristics that specify
the manifold deformation: bending, torsion, rotation, and twist.
(For example, the commutator of the zeroth degree metric form
$\Gamma^\rho_{\mu\nu}$ characterizes the bend, that of the first degree
form $(\Gamma^\rho_{\mu\nu}-\Gamma^\rho_{\nu\mu})$ characterizes the torsion,
the commutator of the third-degree metric form $R^\mu_{\nu\rho\sigma}$
determines the curvature. (For manifolds with closed metric form of first
degree the coefficients of connectedness are symmetric ones.)

Is is evident that the manifolds which are metric ones or possess the
structure have closed metric forms. It is with such manifolds the
exterior differential forms are connected.

If the manifolds are deforming manifolds, this means that their
metric form commutators are nonzero. That is, the metric forms of such
manifolds turn out to be unclosed.

The evolutionary skew-symmetric differential forms are defined on 
manifolds with unclosed metric forms.

Specific properties of evolutionary skew-symmetric differential 
forms and the distinction of evolutionary forms from exterior ones
are connected with the properties of commutators of unclosed metric 
form, which enter into the evolutionary form commutators.

\bigskip
{\bf Distinction between differentials of exterior and evolutionary 
forms.}

The evolutionary differential form of degree $p$ ($p$-form) is written
similarly to exterior differential form. But the evolutionary form
differential cannot be written similarly to that presented for
exterior differential forms. In the evolutionary form
differential there appears an additional term connected with the fact
that the basis of evolutionary form changes. For differential
forms defined on the manifold with unclosed metric form one has
$d(dx^{\alpha_1}\wedge dx^{\alpha_2}\wedge \dots \wedge dx^{\alpha_p})\neq 0$.
(For differential forms defined on the manifold with closed metric form
one has $d(dx^{\alpha_1}\wedge dx^{\alpha_2}\wedge \dots \wedge dx^{\alpha_p})=0$).
For this reason the differential of the evolutionary form $\theta^p$ can be
written as
$$
d\theta^p{=}\!\sum_{\alpha_1\dots\alpha_p}\!da_{\alpha_1\dots\alpha_p}\wedge
dx^{\alpha_1}\wedge dx^{\alpha_2}\dots \wedge
dx^{\alpha_p}{+}\!\sum_{\alpha_1\dots\alpha_p}\!a_{\alpha_1\dots\alpha_p}
d(dx^{\alpha_1}\wedge dx^{\alpha_2}\dots \wedge dx^{\alpha_p})\eqno(9)
$$
where the second term is a differential of unclosed metric form of nonzero
value.

[In further presentation the symbol of summing $\sum$ and the symbol
of exterior multiplication $\wedge$ will be omitted. Summation
over repeated indices will be implied.]

The second term connected with the differential of the basis can be
expressed in terms of the metric form commutator.

For example, let us consider the first-degree form
$\theta=a_\alpha dx^\alpha$. The differential of this form can
be written as
$$d\theta=K_{\alpha\beta}dx^\alpha dx^\beta\eqno(10)$$
where
$K_{\alpha\beta}=a_{\beta;\alpha}-a_{\alpha;\beta}$ are the
components of commutator of the form $\theta$, and
$a_{\beta;\alpha}$, $a_{\alpha;\beta}$ are covariant
derivatives. If we express the covariant derivatives in terms of
the connectedness (if it is possible), they can be written
as $a_{\beta;\alpha}=\partial a_\beta/\partial
x^\alpha+\Gamma^\sigma_{\beta\alpha}a_\sigma$, where the first
term results from differentiating the form coefficients, and the
second term results from differentiating the basis. We arrive at the
following expression for the commutator components of the form $\theta$
$$
K_{\alpha\beta}=\left(\frac{\partial a_\beta}{\partial
x^\alpha}-\frac{\partial a_\alpha}{\partial
x^\beta}\right)+(\Gamma^\sigma_{\beta\alpha}-
\Gamma^\sigma_{\alpha\beta})a_\sigma\eqno(11)
$$
Here the expressions
$(\Gamma^\sigma_{\beta\alpha}-\Gamma^\sigma_{\alpha\beta})$
entered into the second term are just the components of
commutator of the first-degree metric form.

If to substitute the expressions (11) for evolutionary form
commutator into formula (10), we obtain the
following expression for differential of the first degree
skew-symmetric form
$$
d\theta=\left(\frac{\partial a_\beta}{\partial
x^\alpha}-\frac{\partial a_\alpha}{\partial
x^\beta}\right)dx^\alpha dx^\beta+\left((\Gamma^\sigma_{\beta\alpha}-
\Gamma^\sigma_{\alpha\beta})a_\sigma\right)dx^\alpha dx^\beta\eqno(12)
$$
The second term in the expression for differential of skew-symmetric
form is connected with the differential of the manifold metric form,
which is expressed in terms of the metric form commutator.

Thus, the differentials and, correspondingly, the commutators of
exterior and evolutionary forms are of different types. In
contrast to the exterior form commutator, the evolutionary form
commutator includes two terms. These two terms have different
nature, namely, one term is connected with the coefficients of
evolutionary form itself, and the other term is connected with
differential characteristics of manifold. Interaction between
terms of the evolutionary form commutator (interactions between
coefficients of evolutionary form and its basis) provides the
foundation of evolutionary processes that lead to generation of
closed inexact exterior forms.

\bigskip
{\large\bf Properties  of evolutionary forms.}

Above it has been shown that the evolutionary form commutator
includes the commutator of the manifold metric form which is nonzero.
Therefore, the evolutionary form commutator cannot be equal to zero.
This means that the evolutionary form
differential is nonzero. Hence, the evolutionary form, in contrast to
the case of the exterior form, cannot be closed. This leads to
that in the mathematical apparatus of evolutionary forms there arise
new unconventional elements like nonidentical relations and degenerate
transformations. Just such peculiarities allow to describe
evolutionary processes.

Nonidentical relations of the evolutionary form theory, as well as
identical relations of the theory of closed exterior forms,
are relations between the differential and the skew-symmetric form.
In the right-hand side of the identical relation of exterior forms
(see relation (8))  it stands a closed form, which is a differential
as well as the left-hand side. And in the right-hand of the
nonidentical relation of evolutionary form it stands the evolutionary
form that is not closed and, hence, cannot be a differential like
the left-hand side. Such a relation cannot be identical one.

Nonidentical relations are obtained while describing any processes in 
terms of differential equations. (In Appendix 1 we present an example 
of a qualitative investigation of differential equations.) The
relation of such type is obtained, for example, while analyzing the
integrability of the partial differential equation. The equation is
integrable if it can be reduced to the form $d\phi=dU$. However it
appears that, if the equation is not subject to an additional
condition (the integrability condition), the right-hand side turns out
to be an unclosed form and it cannot be expressed as a differential.

Nonidentical relations of evolutionary forms are evolutionary relations
because they include the evolutionary form.
Such nonidentical evolutionary relations appear to be selfvarying
ones.  A variation of any object of the relation in some process leads
to a variation of another object and, in turn, a variation of the latter
leads to a variation of the former. Since one of the objects is
a noninvariant (i.e. unmeasurable) quantity, the other cannot be
compared with the first one, and hence, the process of mutual
variation cannot be completed.

The nonidentity of evolutionary relation is connected with a
nonclosure of the evolutionary form, that is, it is connected with
the fact that the evolutionary form commutator is nonzero. As it has
been pointed out, the evolutionary form commutator includes two terms:
one term specifies the mutual variations of the evolutionary form
coefficients, and the second term (the metric form commutator) specifies
the manifold deformation. These terms have a different nature and cannot
make the commutator vanish. In the process of selfvariation of the
nonidentical evolutionary relation it proceeds an exchange between the
terms of the evolutionary form commutator and this is realized according
to the evolutionary relation. The evolutionary form commutator describes
a quantity that is a moving force of the evolutionary process.

The significance of the evolutionary relation selfvariation consists in
the fact that in such a process it can be realized the conditions of
degenerate transformation under which the closed inexact exterior form
is obtained from the evolutionary form, and from nonidentical relation
the identical relation is obtained.

\bigskip
{\large\bf Analysis of the mathematical
physics equations describing physical processes in material media with the
help of evolutionary forms}

Exterior and evolutionary forms enable one to investigate the integrability
of differential equations. This is due to the fact
that they make it possible to study the conjugacy of the
equations or their derivatives. The type of solutions to differential equations
depends on the conjugacy.  Solutions are invariant if the equations and their derivatives
are conjugated ones. If this is not fulfilled, the solutions prove to be
noninvariant, namely, they are functionals rather then functions.

The qualitative analysis of the equations of mathematical physics with the help of
differential forms shows that any differential equations describing any processes
turn out to be nonintegrable without additional conditions. Additional conditions
under which the equations become integrable can be realized only discretely.
This points to the fact that the solutions of any differential equations of
mathematical physics describing physical processes can be only generalized
(discrete) ones (see Appendix 1). These are precisely generalized
solutions that describe various structures. 

The importance of evolutionary forms consists in the fact that they allow not only
to investigate an integrability of the equations and functional properties of the
solutions. They also allow to describe the process of realization of invariant
solutions in itself and thereby to describe the process of {\it origination }
of physical structures and to disclose the mechanism of processes like
turbulence, generation of waves and vortices, creation of massless particles,
and so on.

Below we will carry out the analysis of the equations of
mathematical physics, which describe physical processes in
material media. These are precisely material media that generate
physical structures making up physical fields. 

It will be shown that the closed exterior forms, which correspond 
to the conservation laws {\it for physical fields} and describe physical 
structures, are obtained from the evolutionary forms that are connected 
with the equations of conservation laws {\it for material 
media}. The conservation laws for material media are balance (differential) 
conservation laws. The process of obtaining closed exterior forms from 
evolutionary ones just describes the process of generating physical structures 
by material media. These conclusions follow from the analysis of the
equations of balance conservation laws with the help of
differential forms.

[Sometimes below it will be used a double notation in subtitles, one in
reference to physical meaning and another in reference to mathematical
meaning.]

\bigskip
{\bf Connection of evolutionary forms with the balance conservation laws.}

{\it The balance conservation laws are those that establish the balance between
the variation of a physical quantity and the corresponding external action.
These are the conservation laws for material systems (material media)} [9].

The balance conservation laws are the conservation laws for energy, linear
momentum, angular momentum, and mass.

The equations of the balance conservation laws are differential (or integral)
equations that describe a variation
of functions corresponding to physical quantities [10-13].
(The specific forms of these equations for thermodynamical and gas
dynamical material systems and the systems of charged particles
will be presented in the Appendix 2).

But it appears that, even without a knowledge of the concrete form
of these equations, with the help of the differential forms one can see
specific features of these equations that elucidate the properties of
the balance conservation laws. To do so it is necessary to study
the conjugacy (consistency) of these equations.

Equations are conjugate if they can be contracted into identical
relations for the differential, i.e. for a closed form.

Let us analyze the equations that describe the balance conservation
laws for energy and linear momentum.

We introduce two frames of reference: the first is an inertial one
(this frame of reference is not connected with the material system), and
the second is an accompanying
one (this system is connected with the manifold built by
the trajectories of the material system elements). The energy equation
in the inertial frame of reference can be reduced to the form:
$$
\frac{D\psi}{Dt}=A
$$
where $D/Dt$ is the total derivative with respect to time, $\psi $ is
the functional of the state that specifies the material system
(every material system has its own functional of the state), $A$ is the
quantity that
depends on specific features of the system and on external energy
actions onto the system. \{The action functional, entropy, wave function
can be regarded as examples of the functional $\psi $. Thus, the
equation for energy presented in terms of the action functional $S$ has
a similar form:
$DS/Dt\,=\,L$, where $\psi \,=\,S$, $A\,=\,L$ is the Lagrange function.
In mechanics of continuous media the equation for
energy of an ideal gas can be presented in the form [13]: $Ds/Dt\,=\,0$, where
$s$ is entropy. In this case $\psi \,=\,s$, $A\,=\,0$. It is worth noting that
the examples presented show that the action functional and entropy play the
same role.\}

In the accompanying frame of reference the total derivative with respect to
time is transformed into the derivative along the trajectory. Equation
of energy is now written in the form
$$
{{\partial \psi }\over {\partial \xi ^1}}\,=\,A_1 \eqno(13)
$$

Here $\psi$  is the functional specifying the state of material system, 
$\xi^1$ is the coordinate along the
trajectory, $A_1$ is the quantity that depends on specific features of
material system and on external (with respect to local domain of material 
system) energy actions onto the system.

In a similar manner, in the accompanying reference system the
equation for linear momentum appears to be reduced to the equation of
the form
$$
{{\partial \psi}\over {\partial \xi^{\nu }}}\,=\,A_{\nu },\quad \nu \,=\,2,\,...\eqno(14)
$$
where $\xi ^{\nu }$ are the coordinates in the direction normal to the
trajectory, $A_{\nu }$ are the quantities that depend on the specific
features of material system and on external force actions.

Eqs. (13) and (14) can be convoluted into the relation
$$
d\psi\,=\,A_{\mu }\,d\xi ^{\mu },\quad (\mu\,=\,1,\,\nu )\eqno(15)
$$
where $d\psi $ is the differential
expression $d\psi\,=\,(\partial \psi /\partial \xi ^{\mu })d\xi ^{\mu }$.

Relation (15) can be written as
$$
d\psi \,=\,\omega \eqno(16)
$$
here $\omega \,=\,A_{\mu }\,d\xi ^{\mu }$ is the skew-symmetrical differential form of the first degree.

Since the balance conservation laws are evolutionary ones, the relation
obtained is also an evolutionary relation.

Relation (16) has been obtained from the equation of the balance conservation
laws for energy and linear momentum. In this relation the form $\omega $
is that of the first degree. If the equations of the balance conservation
laws for angular momentum be added to the equations for energy and linear
momentum, this form in the evolutionary relation will be a form of the
second degree. And in combination with the equation of the balance
conservation law for mass this form will be a form of degree 3.

Thus, in general case the evolutionary relation can be written as
$$
d\psi \,=\,\omega^p \eqno(17)
$$
where the form degree  $p$ takes the values $p\,=\,0,1,2,3$..
(The evolutionary relation for $p\,=\,0$ is similar to that in the differential
forms, and it was obtained from the interaction of energy and time.)

Let us show that relation obtained from the equation
of the balance conservation laws proves to be nonidentical.

To do so we shall analyze relation (16).

In the left-hand side of relation (16) there is a
differential that is a closed form. This form is an invariant
object. The right-hand side of relation (16) involves the differential
form $\omega$ that is not an invariant object because in real processes,
as it will be shown below, this form proves to be unclosed. The commutator of this
form is nonzero. The components of commutator of the form $\omega \,=\,A_{\mu }d\xi ^{\mu }$
can be written as follows:
$$
K_{\alpha \beta }\,=\,\left ({{\partial A_{\beta }}\over {\partial \xi ^{\alpha }}}\,-\,
{{\partial A_{\alpha }}\over {\partial \xi ^{\beta }}}\right )
$$
(here the term connected with the manifold metric form
has not yet been taken into account).

The coefficients $A_{\mu }$ of the form $\omega $ have to be obtained either
from the equation of the balance conservation law for energy or from that for
linear momentum. This means that in the first case the coefficients depend
on the energetic action and in the second case they depend on the force action.
In actual processes energetic and force actions have different nature and appear
to be inconsistent. The commutator of the form $\omega $ consisted of
the derivatives of such coefficients is nonzero.
This means that the differential of the form $\omega $
is nonzero as well. Thus, the form $\omega$ proves to be unclosed and
cannot be a differential like the left-hand side.
This means that relation (16) cannot be an identical one.

In a similar manner one can prove the nonidentity of relation (17).

Hence, without a knowledge of particular expression for the form
$\omega$, one can argue that for actual processes the relation obtained
from the equations corresponding to the balance conservation laws proves to be
nonidentical.

\bigskip
{\bf Physical meaning of nonidentical evolutionary relation.}

The nonidentity of evolutionary relation means that
the balance conservation law equations are inconsistent. And this
indicates that the balance conservation laws are noncommutative. (If the balance
conservation laws be commutative, the equations would be consistent and the
evolutionary relation would be identical).

To what such a noncommutativity of the balance conservation laws leads?

Nonidentical evolutionary relation obtained from the equations of
the balance conservation laws involves the functional that
specifies the material system state. However, since this relation
turns out to be not identical, from this relation one cannot get
the differential $d\psi $  that could point out to the equilibrium
state of material system. The absence of differential means that
the system state is nonequilibrium. That is, due to
noncommutativity of the balance conservation laws the material
system state turns out to be nonequilibrium under effect of
external actions. This points out to the fact that in material
system the internal force acts. (External actions onto local domain of 
material system lead to emergency of internal forces in local domain.) 

The action of internal force leads to a distortion of trajectories of material
system. The manifold made up by the trajectories (the accompanying
manifold) turns out to be a deforming manifold. The differential
form $\omega$,  as well as the forms $\omega^p$ defined on such manifold,
appear to be evolutionary forms. Commutators of these forms will contain
an additional term connected with the commutator of unclosed metric form of
manifold, which specifies the manifold deformation.

\bigskip
{\bf Selfvariation of nonidentical evolutionary relation.
(Selfvariation of nonequilibrium state of material system.)}

The availability of two terms in the commutator of the form $\omega^p $
and the nonidentity of evolutionary relation lead to that the relation
obtained from the balance conservation law equations turns out to be a selfvarying
relation.

Selfvariation of nonidentical evolutionary relation points to the
fact that the nonequilibrium state of material system turns out
to be selfvarying. State of material system changes but remains nonequilibrium
during this process.

It is evident that this selfvariation proceeds under the action of internal force
whose quantity is described by the commutator of the unclosed evolutionary form
$\omega^p $. (If the commutator
be zero, the evolutionary relation would be identical, and this would
point to the equilibrium state, i.e. the absence of internal forces.)
Everything that gives a contribution into the commutator of the form
$\omega^p $ leads to emergency of internal force.

What is the result of such a process of selfvarying the nonequilibrium state
of material system?

\bigskip
{\bf Degenerate transformation. Emergency of closed exterior forms.
(Origination of physical structures.)}

The significance of the evolutionary relation selfvariation consists in
the fact that in such a process it can be realized conditions under
which the closed exterior form is obtained from the evolutionary form.

These are conditions of degenerate transformation. Since the differential
of evolutionary form, which is unclosed, is nonzero, but the differential
of closed exterior form equals zero, the transition from evolutionary
form to closed exterior form is possible only as a degenerate
transformation, namely, a transformation that does not conserve the
differential. And this transition is possible exclusively to closed 
{\it inexact} exterior form, i.e. to the external form being closed on
pseudostructure. The conditions of degenerate transformation are
those of vanishing the commutator (interior one) of the metric form
defining the pseudostructure, in other words, the closure conditions
for dual form.

As it has been already mentioned, the evolutionary differential form
$\omega^p$, involved into nonidentical relation (17) is an unclosed one.
The commutator of this form, and hence the differential, are nonzero.
That is,
$$d\omega^p\ne 0 \eqno(18)$$
If the conditions of degenerate transformation are realized, then from
the unclosed evolutionary form one can obtain the differential form closed
on pseudostructure. The differential of this form equals zero. That is,
it is realized the transition
$d\omega^p\ne 0 \to $ (degenerate transformation) $\to d_\pi \omega^p=0$,
$d_\pi{}^*\omega^p=0$

The relations obtained
$$d_\pi \omega^p=0,  d_\pi{}^*\omega^p=0 \eqno(19)$$
are the closure conditions for exterior inexact form. This means that
it is realized the exterior form closed on pseudostructure.

The realization of closed (on pseudostructure) inexact exterior form
points to emergency of physical structure [14]. 

To the degenerate transformation it must
correspond a vanishing of some functional
expressions, such as Jacobians, determinants, the Poisson
brackets, residues and others. Vanishing these
functional expressions is the closure condition for dual form.

The conditions of degenerate transformation that lead to emergency of
closed inexact exterior form are connected with any symmetries.
Since these conditions are conditions of vanishing the interior
differential of the metric form, i.e. vanishing the interior (rather
than total) metric form commutator, the conditions of degenerate
transformation can be caused by symmetries of coefficients of the metric
form commutator (for example, these can be symmetrical connectednesses).

While describing material system the symmetries can be due to
degrees of freedom of material system.
The translational degrees of freedom, internal degrees
of freedom of the system elements, and so on can be examples of
such degrees of freedom.

And it should be emphasized once more that {\it the degenerate
transformation is realized as a transition from the accompanying
noninertial coordinate system to the locally inertial system}.
The evolutionary form is defined in the noninertial frame of reference
(deforming manifold). But the closed exterior form formatted
is obtained with respect to the locally-inertial frame of reference
(pseudostructure).

The conditions of degenerate transformation (vanishing the dual form commutator) 
 define a pseudostructure. These conditions specify the 
derivative of implicit function, which defines the direction of pseudostructure.

The speeds of various waves are examples of such derivatives:
the speed of light, the speed of sound and of electromagnetic
waves (see the Appendix), the speed of creating particles and so on.

It can be shown that the equations for surfaces of potential
(of simple layer, double layer), equations for one, two, \dots\
eikonals, of the characteristic and of the characteristic surfaces,
the residue equations and so on serve as the equations for
pseudostructures.

The mechanism of creating the pseudostructures lies at the basis of
forming the pseudometric surfaces and their transition into the
metric spaces (see below).

\bigskip
{\bf Obtaining identical relation from nonidentical one.
(Transition of material system into a locally equilibrium
state.)}

On the pseudostructure $\pi$ evolutionary relation (17) converts into
the relation
$$
d_\pi\psi=\omega_\pi^p\eqno(20)
$$
which proves to be an identical relation. Indeed, since the form
$\omega_\pi^p$ is a closed one, on the pseudostructure this form turns
out to be a differential of some differential form. In other words,
this form can be written as $\omega_\pi^p=d_\pi\theta$. Relation (20)
is now written as
$$
d_\pi\psi=d_\pi\theta
$$
There are differentials in the left-hand and right-hand sides of
this relation. This means that the relation is an identical one.

Thus one can see that under degenerate transformation
from evolutionary relation the relation identical on
pseudostructure is obtained. Here it should be emphasized that in this case the
evolutionary relation itself remains to be nonidentical one. The
differential, which equals zero, is an interior one. The evolutionary
form commutator vanishes only on pseudostructure.
The total evolutionary form commutator is nonzero. That
is, under degenerate transformation the evolutionary form differential
vanishes only on pseudostructure. The total differential of the
evolutionary form is nonzero. The evolutionary form remains to be
unclosed.

From relation (20) one can obtain a differential which specifies the
state of material system (and the state function), and this
corresponds to equilibrium state of the system.
But identical relation can be realized only on pseudostructure (which is
specified by the condition of degenerate transformation). This
means that the transition of material system to equilibrium state
proceeds only locally (in the local domain of material system).
In other words, it is realized the transition
of material system from nonequilibrium state to locally
equilibrium one. In this case the global state of material system
remains to be nonequilibrium.

The transition from nonidentical relation (17) obtained from
the balance conservation laws to identical
relation (20) means the following. Firstly, an emergency of the
closed (on pseudostructure) inexact exterior form (right-hand side
of relation (20)) points to an origination of physical structure.
And, secondly, an existence of the state differential (left-hand side
of relation (20))
points to a transition of material system from nonequilibrium state
to the locally-equilibrium state.

Thus one can see that the transition of material system from
nonequilibrium state to locally-equilibrium state is accompanied
by originating differential-geometrical structures, which are
physical structures.

The emergency of physical structures in the
evolutionary process reveals in material system as an emergency of
certain observable formations, which develop spontaneously. Such
formations and their manifestations are fluctuations, turbulent
pulsations, waves, vortices, creating massless particles, and others.
The intensity of such formations is controlled by a quantity
accumulated by the evolutionary form commutator at the instant in
time of originating physical structures.

Here the following should be pointed out. Physical structures are
generated by local domains of material system. They are elementary
physical structures. By combining with one another they can form
large-scale structures and physical fields.

\bigskip
The availability of physical structures points out to fulfilment
of conservation laws. These are conservation laws for physical fields.
The process of generating physical structures (forming physical
fields) demonstrates a connection of these conservation laws, which
had been named as exact ones, with the balance (differential)
conservation laws for material media.
The physical structures that correspond to the exact
conservation laws are produced by material system in the evolutionary
processes, which are based on the interaction of noncommutative
balance conservation laws.

{\it Noncommutativity of balance conservation laws for material media
and their controlling role in evolutionary processes
accompanied by emerging  physical structures practically
have not been taken into account in the explicit form anywhere}. The
mathematical apparatus of evolutionary differential forms enables one
to take into account and to describe these points [9].

\bigskip
{\bf Physical meaning of the duality of closed exterior forms as
conservative quantities and as potential forces. (Potential forces)}

The duality of closed exterior forms as conservative quantities and
as potential forces points to that an unmeasurable quantity, which is
described by the evolutionary form commutator (recall, that all
external, with respect to local domain, actions make a contribution into
this commutator) and acts as an internal force,
is converted into a measurable quantity that acts as a potential force.

Where, from what, and on what the potential force acts?

The potential force is an action of created (quantum) formation onto
the local domains of the material system over which it is translated.
And if the internal force acts in the interior of the local domain of
material system (and it caused that to deform), the potential force
acts onto the neighboring domain. The local domain gets rid of its
internal force and modifies that into a potential force which acts onto
neighboring domains. An unmeasurable quantity, that acts in local
domain as an internal force, is transformed into a measurable quantity
of the observable formation (and the physical structure as well) that
is emitted from the local domain and acts onto neighboring domain
as a force equal to this quantity.

If the external actions equal zero (the evolutionary form commutator be
equal to zero), then internal and potential forces equal zero.

Thus, one has to distinguish the forces of three types: 1)
external forces (the actions being external with respect to local
domain), 2) internal forces that originate in local domains of
material system due to the fact that the physical quantities of
material system changed by external actions turn out to be
inconsistent, and 3) the potential forces are forces of the action
of the formations (corresponding to physical structures) onto
material system.

The potential force, whose value is conditioned by the
quantity of the commutator
of the evolutionary form $\omega^p$ at the instant of the formation
production, acts normally to the  pseudostructure, i.e. with
respect to
the integrating direction, along which the interior differential
(the closed form) is formed.
The potential forces are described, for example,
by jumps of derivatives in the direction normal to
characteristics, to potential surfaces and so on.
This corresponds to the fact that the evolutionary form
commutators along these directions are nonzero.

The duality  of closed inexact form as a conservative
quantity and as a potential force shows that
potential forces are the action of formations corresponding to
physical structures onto material system.

\bigskip
{\bf Connection of the characteristics of the structures originated with the
characteristics of differential forms.
(Characteristics of physical structures and the formation created)}

Since the closed inexact exterior form corresponding to physical
structure was obtained from the evolutionary form, it is evident that the
characteristics of physical structure originated has to be connected with
those of the evolutionary form and of the manifold on which this form is
defined as well as with the conditions of degenerate transformation and with
the values of commutators of the evolutionary form and the manifold metric form.

The conditions of degenerate transformation, i.e. symmetries caused
by degrees of freedom of material system, determine the equation for
pseudostructures.

The closed exterior forms corresponding to physical structures are
conservative quantities. These conservative quantities describe
certain charges.

The first term of the commutator of evolutionary form determines
the value of discrete change (the quantum),
which the quantity conserved on the pseudostructure undergoes during
transition from one pseudostructure to another. The second term of the
evolutionary form commutator specifies a characteristics that fixes the
character of the manifold deformation, which took place before
physical structure emerged.  (Spin is an example of such
a characteristics).

Characteristics of physical structures depends in addition on the properties of
material system generating these structures.

The closed exterior forms obtained correspond to the state differential
for material system. The differentials of entropy, action, potential
and others are examples of such differentials.

As it was already mentioned, in material system the created physical structure
is revealed as an observable formation. It is evident that the characteristics
of the formation, as well as those of created physical structure, are
determined by the evolutionary form and its commutator and by the material system
characteristics.

It can be shown that the following correspondence between characteristics
of the formations emerged and characteristics of evolutionary forms, of the
evolutionary form commutators and of material system is established:

1) an intensity of the formation (a potential force)
$\leftrightarrow$ {\it the  value of the first term in the
commutator of evolutionary form} at the instant when the formation
is created;

2) vorticity (an analog of spin)  $\leftrightarrow$ {\it the second term in the commutator
that is connected with the metric form commutator};

3) an absolute speed of propagation of created formation (the
speed in the inertial frame of reference) $\leftrightarrow$ {\it additional
conditions connected with degrees of freedom of material system};

4) a speed of the formation propagation relative to material system
$\leftrightarrow $  {\it additional
conditions connected with degrees of freedom of material system plus
the velocity of elements of local domain}.

\bigskip
{\bf Parameters of differential forms.
(Classification of physical structures)}

The connection of physics structures with
skew-symmetric differential forms allows to introduce a classification
of these structures in dependence on parameters that specify
skew-symmetric differential forms and enter into nonidentical and
identical relation. To determine these parameters one has to consider
the problem of integration of nonidentical evolutionary relation.

Since the identical relation obtained from nonidentical
evolutionary relation contains only differential and the closed
form also is a differential, one can integrate (on
pseudostructure) this relation and obtain a relation with the
differential forms of less by one degree. From the relation
obtained, which will be nonidentical one, under degenerate
transformation it can be obtained  new identical relation that can
be integrated once more.

Thus, from the nonidentical relation, which contains the evolutionary
form of degrees $p$, it can be obtained identical relations with closed
inexact forms of degrees $k$, where $k$ ranges from $p$ to $0$. That is,
evolutionary forms of degree $p$ can generate closed inexact
forms of degrees $k=p$, $k=p-1$, \dots, $k=0$. Under degenerate conditions
from closed inexact forms of zero degree it is obtained an exact exterior form
of zero degree which the metric structure corresponds to.

In addition to these parameters, another parameter appears, namely, the
dimension of space $n$ If the evolutionary relation generates the closed
forms of degrees $k=p$, $k=p-1$, \dots, $k=0$, to them there correspond
the pseudostructures of dimensions $(n+1-k)$.

The parameters of evolutionary and exterior forms that
follow from the evolutionary forms allow to introduce a classification
of physical structures that defines a type of physical structures
and, accordingly, of physical fields and interactions.

The type of physical structures (and,
accordingly, of physical fields) generated by the evolutionary
relation depends on the degree of differential forms $p$ and $k$
and on the dimension of original inertial space $n$. 
(Here $p$ is the degree of evolutionary form in nonidentical 
relation that is connected with a number of interacting balance
conservation laws, and $k$ is the  degree of closed form generated by 
nonidentical relation. Recall that the interaction of balance 
conservation laws for energy and linear momentum corresponds to the 
value $p=1$, with the balance conservation law for angular momentum in 
addition this corresponds to the value $p=2$, and with the balance 
conservation law for mass in addition it corresponds to the value 
$p=3$. The value $p=0$ corresponds to interaction between time and the 
balance conservation law for energy.)

By introducing a
classification by numbers $p$, $k$, $n$ one can understand the
internal connection between various physical fields. Since physical
fields are the carriers of interactions, such classification enables
one to see a connection between interactions.

On the basis of the properties of evolutionary forms that correspond to
the conservation laws, one can suppose that such a classification may be
presented in the form of the table given below.
This table corresponds to elementary particles.

\{It should be emphasized the following. Here the concept of ``interaction"
is used in a twofold meaning: an interaction of the balance conservation laws
that relates to material systems, and the physical concept of ``interaction"
that relates to physical fields and reflects the interactions of physical
structures, namely, it is connected with exact conservation laws\}.

\bigskip

\centerline{TABLE}

\noindent
\begin{tabular}{@{~}c@{~}c@{~}c@{~}c@{~}c@{~}c@{~}}
\bf interaction&$k\backslash p,n$&\bf 0&\bf 1&\bf 2&\bf 3

\\
\hline
\hline
\bf gravitation&\bf 3&&&&
    \begin{tabular}{c}
    \bf graviton\\
    $\Uparrow$\\
    electron\\
    proton\\
    neutron\\
    photon
    \end{tabular}

\\
\hline
    \begin{tabular}{l}
    \bf electro-\\
    \bf magnetic
    \end{tabular}
&\bf 2&&&
    \begin{tabular}{c}
        \bf photon2\\
    $\Uparrow$\\
    electron\\
    proton\\
    neutrino
    \end{tabular}
&\bf photon3

\\
\hline
\bf weak&\bf 1&&
    \begin{tabular}{c}
    \bf neutrino1\\
    $\Uparrow$\\
    electron\\
    quanta
    \end{tabular}
&\bf neutrino2&\bf neutrino3

\\
\hline
\bf strong&\bf 0&
    \begin{tabular}{c}
    \bf quanta0\\
    $\Uparrow$\\
    quarks?
    \end{tabular}
&
    \begin{tabular}{c}
    \bf quanta1\\
    \\

    \end{tabular}
&
\bf quanta2&\bf quanta3

    \\
\hline
\hline
    \begin{tabular}{c}
    \bf particles\\
    material\\
    nucleons?
    \end{tabular}
&
    \begin{tabular}{c}
    exact\\
    forms
    \end{tabular}
&\bf electron&\bf proton&\bf neutron&\bf deuteron?
\\
\hline
N&&1&2&3&4\\
&&time&time+&time+&time+\\
&&&1 coord.&2 coord.&3 coord.\\
\end{tabular}

In the Table the names of the particles created are given. Numbers placed                                             
near particle names correspond to the space dimension. Under the names of 
particles the
sources of interactions are presented. In the next to the last row we
present particles with mass (the elements of material system) formed by
interactions (the exact forms of zero degree obtained by sequential 
integrating the evolutionary relations with the evolutionary forms of 
degree $p$ corresponding to these particles). In the bottom row the 
dimension of the {\it metric} structure created is presented.

From the Table one can see the correspondence between the degree $k$ of 
the closed forms realized and the type of interactions. Thus, $k=0$ 
corresponds to strong interaction, $k=1$ corresponds to weak interaction,
$k=2$ corresponds to electromagnetic interaction, and $k=3$ corresponds
to gravitational interaction.
The degree $k$ of the closed forms realized and the number of interacting
balance conservation laws determine the type of interactions and the type
of particles created. The properties of particles are governed by the 
space dimension. The last property is connected with the fact that
closed forms of equal degrees $k$, but obtained from the evolutionary
relations acting in spaces of different dimensions $n$, are distinctive
because they are defined on pseudostructures of different dimensions
(the dimension of pseudostructure $(n+1-k)$ depends on the dimension
of initial space $n$). For this reason the realized physical structures
with closed forms of equal degrees $k$ are distinctive in their 
properties.

The parameters $p$, $k$, $n$ can range from 0 to 3. They determine some
completed cycle. The cycle involves four levels, to each of which there
correspond their own values of $p$ ($p=0,1,2,3$) and space
dimension $n$. 

In the Table one cycle of forming physical structures is presented.
Each material system has his own completed cycle. This distinguishes one 
material system from another system. One completed cycle can serve as 
the beginning of another cycle (the structures formed in the preceding 
cycle serve as the sources of interactions for beginning a new cycle). 
This may mean that one material system (medium) proves to be imbedded 
into the other material system (medium). The sequential cycles 
reflect the properties of sequentially imbedded material systems.
And yet a given level has specific 
properties that are inherent characteristics of the same level in
another cycles. This can be seen, for example, from comparison of
the cycle described and the cycle in which to the exact forms
there correspond conductors, semiconductors, dielectrics, and
neutral elements. The properties of elements of the third
level, namely, of neutrons in one cycle and of dielectrics in
another, are identical to the properties of so called "magnetic
monopole" [15,16].

\bigskip
{\bf Forming pseudometric and metric spaces}

The mechanism of creating the pseudostructures lies at the basis of
forming the pseudometric surfaces and their transition into metric
spaces [17]. (It should be pointed out that the eigenvalues and
the coupling constants appear as the conjugacy conditions for
exterior or dual forms, the numerical constants are the conjugacy
conditions for exact forms.)

It was shown above that the evolutionary relation of degree $p$ can
generate (in the presence  of degenerate transformations) closed forms
of the degrees  $p, p-1,.., 0$. While generating closed forms of
sequential degrees  $k=p, k=p-1,.., k=0$ the pseudostructures of
dimensions $(n+1-k)$: $1, ..., n+1$ are obtained. As a result of
transition to the exact closed form of zero degree the metric structure
of the dimension $n+1$ is obtained. Under influence of external
action (and in the presence of degrees of freedom) the material system
can transfer the initial inertial space into the space of the dimension
$n+1$.

Sections of the cotangent bundles (Yang-Mills fields),
cohomologies by de Rham, singular cohomologies, pseudo-Riemannian and
pseudo-Euclidean spaces, and others are examples of psedustructures
and spaces that are formed by pseudostructures. Euclidean and Riemannian
spaces are examples of metric manifolds that are obtained when going
to the exact forms.

What can be said about the pseudo-Riemannian manifold and Riemannian space?

The distinctive property of the Riemannian manifold is an availability of
the curvature. This means that the metric form commutator of the third
degree is nonzero. Hence, the  commutator of the
evolutionary form of third degree ($p=3$), which involves into
itself the metric form commutator, is not equal to zero. That is,
the evolutionary form that enters into the evolutionary relation is unclosed,
and the relation is nonidentical one.

When realizing pseudostructures of the dimensions  $1, 2, 3, 4$
and obtaining the closed inexact forms of the degrees $k=3, k=2,
k=1, k=0$ the pseudo-Riemannian space is formed, and the
transition to the exact form of zero degree corresponds to the
transition to the Riemannian space.

It is well known that while obtaining the Einstein equations it was
suggested that there are fulfilled the conditions [6,18]: 1) the Bianchi 
identity is satisfied, 2) the coefficients of connectedness are symmetric,
3) the condition that the coefficients of connectedness are the Christoffel
symbols, and 4) an existence of the transformation under which the
coefficients of connectedness vanish. These conditions are the
conditions of realization of degenerate transformations for
nonidentical relations obtained from the evolutionary nonidentical
relation  with evolutionary form of the degree $p=3$ and
after going to the identical relations. In this case to the Einstein
equation the identical relations with forms of the first degree are 
assigned.

\bigskip
\centerline {\large\bf Conclusion}

Results of the analysis carried out have shown the following.

Invariant and covariant properties of closed exterior and dual
forms, which correspond to the conservation laws for physical fields,
make up the foundations of field
theories. The field theories operators are built on the basis of
gauge transformations of closed exterior forms.
Properties of closed exterior and dual forms explicitly or implicitly
manifest themselves essentially in all formalisms of field theories.
The degrees of closed exterior forms establish the classification of 
physical fields and interactions, and this discloses an internal 
connection between various physical fields and a common basis of 
corresponding field theories. This shows that the theory of closed 
exterior forms can be useful in establishing the unified field theory.

Evolutionary forms, which are obtained from the equations of balance 
conservation laws for material media, answer
the question of how are realized the closed exterior forms that
correspond to field theories.

This explains the process of originating physical fields and gives the
answer to many questions of field theories.

Firstly, this shows that physical fields are generated by material media. 
The conservation laws for material media, i.e. the balance conservation 
laws for energy, linear momentum, angular momentum, and mass, which 
are noncommutative ones, play a controlling role in these processes. 
This is precisely the noncommutativity of the balance conservation laws 
produced by external actions onto material system, which is a moving 
force of evolutionary processes leading to emergency of physical 
structures (to which exact conservation laws are assigned).
And thus the causality of physical processes and phenomena is explained. 
Since physical fields are made up by discrete physical structures, 
this points to a quantum character of field theories.

Secondly, it becomes clear a connection of field theory with the 
equations of mathematical physics describing physical processes in 
material media. The postulates, which field theories are built on, are 
the closure conditions for exterior forms obtained from the evolutionary 
forms connected with these equations.

The connection between closed exterior forms corresponding to field
theories and the evolutionary forms obtained from the equations for
material media discloses a meaning of the field theory parameters.
They relate to the number ($p$) of interacting noncommutative balance
conservation laws and to the degrees ($k$) of closed exterior forms
realized. Hence it arises a possibility to classify physical
fields and interactions according to the parameters $p$ and $k$.

\bigskip
The results obtained on the basis of the theory of skew-symmetric differential
forms do not contradict to any physical theories. And yet the methodical results
of this theory enable one to understand internal connections between physical fields,
between physical fields and material media, between field theories and the equations
of mathematical physics, to understand a mechanism of emergency of physical structures
and the causality of physical processes and phenomena.

The theory of skew-symmetric differential forms, which unites
the theory of closed exterior forms constituting the basis of field theories
and the theory of evolutionary forms generating closed inexact
exterior forms, can serve as
an approach to the general field theory. Such a theory enables one
not only to describe physical fields, but also shows how the
physical fields are produced, what generates them, and what is a cause
of these processes.

\bigskip
Below in Appendices the example of qualitative investigation of the solutions to 
differential equations and the analysis of the principles of
thermodynamics, the equations for gas dynamic system and the equations 
of electromagnetic field with the help of skew-symmetric differential 
forms are presented.

\bigskip
\centerline {\large\bf Acknowledgments} 

The author made reports at the seminar of the Institute of General Physics of 
Russian Academy of Sceince many times. She thanks the head of the seminar Prof. 
Anry Rukhadze and the participants for useful discussions and comments. 

The author also thanks the orgenazers of conferences on gravitation, 
theory of relativity, nonlinear acoustics, turbulence, the interacions of 
elementary particles, symmetries, and algebra and geometry for invitations, 
hospitality and helpful discussions. 

I am also thankful to Prof. R.Kiehn for his attention to my works and for multiple 
stimulating discussions. 

\bigskip
\rightline{\large\bf Appendix 1}

\centerline {\large\bf Qualitative investigation of the solutions }
\centerline {\large\bf of differential equations}

The presented method of investigating the solutions to differential
equations is not new. Such an approach was developed by Cartan [19] in
his analysis of the integrability of differential equations.
Here this approach is outlined to demonstrate the role of
skew-symmetric differential forms.

The basic idea of the qualitative investigation of the solutions to
differential equations can be clarified by the example of the
first-order partial differential equation.

Let
$$ F(x^i,\,u,\,p_i)=0,\quad p_i\,=\,\partial u/\partial x^i \eqno(A1.1)$$
be the partial differential equation of the first order. Let us consider
the functional relation
$$ du\,=\,\theta\eqno(A1.2)$$
where $\theta\,=\,p_i\,dx^i$ (the summation over repeated indices is
implied). Here $\theta\,=\,p_i\,dx^i$ is the differential form of the
first degree.

The specific feature of functional relation (A1.2) is that in
the general case this relation turns out to be nonidentical.

The left-hand side of this relation involves a differential, and
the right-hand side includes the differential form
$\theta\,=\,p_i\,dx^i$. For this relation to be identical, the
differential form $\theta\,=\,p_i\,dx^i$ must be a differential as well
(like the left-hand side of relation (A1.2)), that is, it has to be a
closed exterior differential form. To do this it requires the commutator
$K_{ij}=\partial p_j/\partial x^i-\partial p_i/\partial x^j$ of the
differential form $\theta $ has to vanish.

In general case, from equation (A1.1) it does not follow (explicitly)
that the derivatives $p_i\,=\,\partial u/\partial x^i $, which obey
to the equation (and given boundary or initial conditions of the
problem), make up a differential. Without any supplementary conditions
the commutator of the differential form $\theta $ defined as
$K_{ij}=\partial p_j/\partial x^i-\partial p_i/\partial x^j$ is
not equal to zero. The form $\theta\,=\,p_i\,dx^i$ occurs to be
unclosed and is not a differential like the left-hand side of relation
(A1.2). Functional relation (A1.2) appears to be nonidentical: the
left-hand side of this relation is a differential, but the right-hand
side is not a differential.  (The skew-symmetric differential
form $\theta\,=\,p_i\,dx^i$, which enters into functional relation (A1.2),
is the example of evolutionary skew-symmetric differential forms.)

The nonidentity of functional relation (A1.2) points to a fact
that without additional conditions derivatives of the initial
equation do not make up a differential. This means that the
corresponding solution to the differential equation $u$ will not be
a function of $x^i$. It will depend on the commutator of the form
$\theta $, that is, it will be a functional.

To obtain the solution that is a function (i.e., derivatives of this
solution form a differential), it is necessary to add the closure
condition for the form $\theta\,=\,p_idx^i$ and for the dual form
(in the present case the functional $F$ plays a role of the form
dual to $\theta $):
$$\cases {dF(x^i,\,u,\,p_i)\,=\,0\cr
d(p_i\,dx^i)\,=\,0\cr}\eqno(A1.3)$$
If we expand the differentials, we get a set of homogeneous equations
with respect to $dx^i$ and $dp_i$ (in the $2n$-dimensional
space -- initial and tangential):
$$\cases {\displaystyle \left ({{\partial F}\over {\partial x^i}}\,+\,
{{\partial F}\over {\partial u}}\,p_i\right )\,dx^i\,+\,
{{\partial F}\over {\partial p_i}}\,dp_i \,=\,0\cr
dp_i\,dx^i\,-\,dx^i\,dp_i\,=\,0\cr} \eqno(A1.4)$$
The solvability conditions for this system (vanishing of the determinant
composed of coefficients at $dx^i$, $dp_i$) have the form:
$$
{{dx^i}\over {\partial F/\partial p_i}}\,=\,{{-dp_i}\over 
{\partial F/\partial x^i+p_i\partial F/\partial u}} \eqno(A1.5) 
$$
These conditions determine an integrating direction, namely, a
pseudostructure, on which the form $\theta \,=\,p_i\,dx^i$ occurs to be
closed one, i.e. it becomes a differential, and from relation (A1.2) the
identical relation is produced. If conditions (A1.5), that may be called
the integrability conditions, are satisfied, the derivatives constitute
a differential $\delta u\,=\,p_idx^i\,=\,du$ (on the pseudostructure),
and the solution becomes a function.
Just such solutions, namely, functions on pseudostructures
formed by the integrating directions, are so-called generalized
solutions. The derivatives of the generalized solution constitute
the exterior form, which is closed on pseudostructure.

Since the functions that are generalized solutions
are defined only on pseudostructures, they have discontinuities in
derivatives in the directions being transverse to pseudostructures.
The order of derivatives with discontinuities
is equal to  the exterior form degree. If the form of zero
degree is involved in the functional relation, the function itself,
being a generalized solution, will have discontinuities.

If we find the characteristics of equation (A1.1), it appears that
conditions (A1.5) are equations for characteristics [20].
That is, the characteristics are examples of
the pseudostructures on which the derivatives of differential equation                                       
constitute closed forms and the solutions turn out to be functions
(generalized solutions).

Here it is worth noting that coordinates of the equations for
characteristics are not identical to independent coordinates of
the initial space on which equation (A1.1) is defined. The transition 
from initial space to characteristic manifold appears to be a
{\it degenerate} transformation, namely, the determinant of the system
of equations (A1.4) becomes zero. The derivatives of  equation (A1.1) are
transformed from tangent space to cotangent one.
The transition from the tangent space, where the commutator of the form
$\theta$ is nonzero (the form is unclosed, the derivatives do not form a
differential), to the characteristic manifold, namely, the cotangent
space, where the commutator becomes equal to zero (a closed exterior
form is formed, i.e. the derivatives make up a differential), is the
example of degenerate transformation.

The solutions to all differential equations have similar functional properties.
And, if the order of differential equation is $k$, the
functional relation with $k$-degree form corresponds to this
equation. For ordinary differential equations the commutator is produced
at the expense of the conjugacy of derivatives of the functions desired
and those of the initial data (the dependence of the solution on the
initial data is governed by the commutator).

In a similar manner one can also investigate  the solutions to a system
of partial differential equations and the solutions to ordinary
differential equations (for which the nonconjugacy of desired functions
and initial conditions is examined).

It can be shown that  the solutions to equations of mathematical
physics, on which no additional external conditions are imposed, are
functionals. The solutions prove to be exact only under realization of
additional requirements, namely, the conditions of degenerate
transformations like vanishing determinants, Jacobians and so on, that
define integral surfaces. Characteristic manifolds, the
envelopes of characteristics, singular points, potentials of simple and
double layers, residues and others are the examples of such surfaces.

The dependence of the solution on the commutator can lead to instability
of the solution. Equations that do not provided with the integrability
conditions (the conditions such as, for example, the characteristics,
singular points, integrating factors and others) may have unstable
solutions. Unstable solutions appear in the case when additional
conditions are not realized and no exact solutions (their derivatives
form a differential) are formed. Thus, the solutions to the equations
of elliptic type can be unstable.

Investigation of nonidentical functional relations lies at the basis
of the qualitative theory of differential equations. It is well known
that the qualitative theory of differential equations is based on the
analysis of unstable solutions and integrability conditions. From the
functional relation it follows that the dependence of the solution on
the commutator leads to instability, and the closure conditions of the
forms constructed by derivatives are integrability conditions. One
can see that the problem of unstable solutions and integrability
conditions appears, in fact, to be reduced to the question  of under
what conditions the identical relation for closed form is produced
from the nonidentical relation that corresponds to the relevant
differential equation (the relation such as (A1.2)), the identical relation
for closed form is produced. In other words, whether or not the
solutions are functionals? This is, the
analysis of the correctness of setting the problems of mathematical
physics is reduced to the same question.

Here the following should be emphasized. When the degenerate
transformation from the initial nonidentical functional relation is
performed, an integrable identical relation is obtained. As the result
of integrating, one obtains a relation that contains exterior forms of
less by one degree and which once again proves to be (in the general
case without additional conditions) nonidentical. By integrating the
functional relations obtained sequentially  (it is possible only under
realization of the degenerate transformations) from the initial
functional relation of degree $k$ one can obtain $(k+1)$ functional
relations each involving exterior forms of one of degrees:
$k, \,k-1, \,...0$. In particular, for the first-order partial
differential equation it is also necessary to analyze the functional
relation of zero degree.

Thus, application of skew-symmetric differential forms allows one
to reveal the functional properties of the solutions to differential
equations.

\bigskip
{\bf Analysis of field equations}

Field theory is based on the conservation laws. The conservation laws
are described by the closure conditions of the exterior differential
forms. It is evident that the solutions to the equations of field
theory describing physical fields can be only generalized solutions,
which correspond to closed exterior differential forms.
The generalized solutions can have a differential
equation, which is subject to the additional conditions.

Let us consider what equations are obtained in this case.

Return to equation (A1.1).
Assume that the  equation does not explicitly depend on $u$ and is
resolved with respect to some variable, for example $t$, that is, it
has the form of
$${{\partial u}\over {\partial t}}\,+\,E(t,\,x^j,\,p_j)\,=\,0, \quad p_j\,=
\,{{\partial u}\over {\partial x^j}}\eqno(A1.6)
$$
Then integrability conditions (A1.5) (the closure conditions of the
differential form $\theta =p_idx^i$  and the corresponding dual form)                                   
can be written as (in this case $\partial F/\partial p_1=1$)
$${{dx^j}\over {dt}}\,=\,{{\partial E}\over {\partial p_j}}, \quad
{{dp_j}\over {dt}}\,=\,-{{\partial E}\over {\partial x^j}}\eqno(A1.7)$$

These are the characteristic relations for equation (A1.6). As it is well
known, the canonical relations have just such a form.

As a result we conclude that the canonical relations are the
characteristics of equation (A1.6) and the integrability conditions for
this equation.

The canonical relations obtained from the closure condition of the
differential form $\theta = p_idx^i$ and the corresponding dual form,
are the examples of the identical relation of the theory of exterior
differential forms.

Equation (A1.6) provided with the supplementary conditions, namely, the
canonical relations (A1.7), is called the Hamilton-Jacobi equation [20].
In other words, the equation whose derivatives obey the canonical
relation is referred to as the Hamilton-Jacobi equation. The derivatives
of this equation form the differential, i.e. the closed exterior
differential form:
$\delta u\,=\,(\partial u/\partial t)\,dt+p_j\,dx^j\,=\,-E\,dt+p_j\,dx^j\,=\,du$.

The equations of field theory belong to this type.
$${{\partial s}\over {\partial t}}+H \left(t,\,q_j,\,{{\partial s}\over {\partial q_j}}
\right )\,=\,0,\quad
{{\partial s}\over {\partial q_j}}\,=\,p_j \eqno(A1.8)$$
where $s$ is the field function for the action functional
$S\,=\,\int\,L\,dt$. Here $L$ is the Lagrange function, $H$ is the
Hamilton function:
$H(t,\,q_j,\,p_j)\,=\,p_j\,\dot q_j-L$, $p_j\,=\,\partial L/\partial \dot q_j$.
The closed form $ds\,=-\,H\,dt\,+\,p_j\,dq_j$ (the Poincare invariant)
corresponds to equation (A1.8).

A peculiarity of the degenerate transformation can be considered by the
example of the field equation.

Here the degenerate transformation is a transition from the Lagrange
function to the Hamilton function. The equation for the Lagrange
function, that is the Euler variational equation, was obtained from the
condition $\delta S\,=\,0$, where $S$ is the action functional. In the
real case, when forces are nonpotential or couplings are nonholonomic,
the quantity $\delta S$ is not a closed form, that is,
$d\,\delta S\,\neq \,0$. But the Hamilton function is obtained from the
condition $d\,\delta S\,=\,0$ which is the closure condition for the
form $\delta S$. The transition from the Lagrange function $L$ to the
Hamilton function $H$ (the transition from variables $q_j,\,\dot q_j$
to variables $q_j,\,p_j=\partial L/\partial \dot q_j$) is a transition
from the tangent space, where the form is unclosed, to the cotangent
space with a closed form. This transition is
a degenerate one.

The invariant field theories used only nondegenerate transformations
that conserve a differential. There exists a relation between nondegenerate
transformations and degenerate transformations. In the case under consideration
the degenerate transformation is a transition from the tangent space
($q_j,\,\dot q_j)$) to the cotangent (characteristic) manifold
($q_j,\,p_j$),  but the nondegenerate transformation is a
transition from one characteristic manifold ($q_j,\,p_j$) to another
characteristic manifold ($Q_j,\,P_j$). \{The formula of canonical
transformation can be written as $p_jdq_j=P_jdQ_j+dW$, where $W$ is the
generating function\}.

\bigskip
\rightline{\large\bf Appendix 2}

\centerline {\large\bf The analysis of balance conservation laws for }
\centerline {\large\bf thermodynamic and gas dynamic systems and for}
\centerline {\large\bf the system of charged particles}

\subsection*{Thermodynamic systems}

The thermodynamics is based on the first and second principles of thermodynamics
that were introduced as postulates [21].
The first principle of thermodynamics, which can be written in the form
$$dE\,+\,dw\,=\,\delta Q\eqno(A2.1)$$
follows from the balance conservation laws for energy and linear momentum
(but not only from the conservation law for energy). This is analogous to
the evolutionary relation for the thermodynamic system. Since $\delta Q$
is not a differential, relation (A2.1) which corresponds to the first principle
of thermodynamics, as well as the evolutionary relation, appears to be a
nonidentical relation.
This points to a noncommutativity of the balance conservation
laws (for energy and linear momentum) and to a nonequilibrium state of the
thermodynamic system.

If condition of the integrability be satisfied, from the nonidentical
evolutionary relation, which corresponds to the first principle of
thermodynamics, it follows an identical relation. It is an identical relation
that corresponds to the second principle of thermodynamics.

If $dw\,=\,p\,dV$,  there is the integrating factor
$\theta$ (a quantity which depends only on the characteristics of the system),
where $1/\theta\,=\,pV/R$ is called the temperature $T$ [21].
In this case the form $(dE\,+\,p\,dV)/T$ turns out to be a differential
(interior) of some quantity that referred to as entropy $S$:
$$(dE\,+\,p\,dV)/T\,=\,dS \eqno(A2.2)$$

If the integrating factor $\theta=1/T$ has been
realized, that is, relation (A2.2) proves to be satisfied, from relation
(A2.1),
which corresponds to the first principle of thermodynamics,
it follows
$$dS\,=\,\delta Q/T \eqno(A2.3)$$
This is just the second principle of thermodynamics for reversible processes.
This takes place when the heat input is the only action onto the system.

If in addition to the heat input the system experiences a certain mechanical
action
(for example, an influence of boundaries), we obtain
$$dS\, >\,\delta Q/T \eqno (A2.4)$$
that corresponds to the second principle of thermodynamics for irreversible
processes.

In the case examined above the differential of entropy (rather than entropy
itself) becomes a closed form. $\{$In this case entropy  manifests itself
as the thermodynamic potential, namely, the function of state. To the
pseudostructure there corresponds
the state equation that determines the temperature dependence on the
thermodynamic variables$\}$.

For entropy to be a closed form itself,  one more
condition must be realized. Such a condition could be the realization of the
integrating direction, an example of that is the speed of sound:
$a^2\,=\,\partial p/\partial \rho\,=\,\gamma\,p/\rho$. In this case it is valid
the equality $ds\,=\,d(p/\rho ^{\lambda })\,=\,0$ from which it follows that
entropy $s\,=\,p/\rho ^{\lambda }\,=\hbox{const}$ is a closed form (of zero degree).
$\{$However it does not mean that a state of the gaseous system is identically
isoentropic. Entropy is constant only along the integrating direction (for
example, on the adiabatic curve or on the front of the sound wave), whereas in
the direction normal to the integrating direction the normal derivative of
entropy has a break$\}$.

\subsection*{Gas dynamical systems}

We take the simplest gas dynamical system, namely, a flow of ideal
(inviscous, heat nonconductive) gas [13].

Assume that the gas (the element of gas dynamic system) is a thermodynamic
system in the state of local equilibrium (whenever the gas dynamic system
itself may be
in nonequilibrium state), that is, it is satisfied the relation [21]
$$Tds\,=\,de\,+\,pdV \eqno(A2.5)$$
where $T$, $p$ and $V$ are the temperature, the pressure and the gas
volume, $s$ and $e$ are entropy and internal energy per unit volume.

Let us introduce two frames of reference: an inertial one that is not connected
with material system and an accompanying frame of reference that is connected
with the manifold formed by the trajectories of the material system elements.

The equation of the balance conservation law of energy for ideal gas can
be written as [13]
$${{Dh}\over {Dt}}- {1\over {\rho }}{{Dp}\over {Dt}}\,=\,0 \eqno(A2.6)$$
where $D/Dt$ is the total derivative with respect to time (if to denote
the spatial coordinates by $x_i$ and the velocity components by $u_i$,
$D/Dt\,=(\,\partial /\partial t+u_i\partial /\partial x_i$). Here  $\rho=1/V $
and $h$ are respectively the mass and the entalpy densities of the gas.

Expressing entalpy in terms of internal energy $e$ using the formula
$h\,=\,e\,+\,p/\rho $ and using relation (A2.5), the balance conservation law
equation (A2.6) can be put to the form
$${{Ds}\over {Dt}}\,=\,0 \eqno(A2.7)$$

And respectively, the equation of the balance conservation law for linear
momentum can be presented as [13,22]
$$\hbox {grad} \,s\,=\,(\hbox {grad} \,h_0\,+\,{\bf U}\times \hbox {rot} {\bf U}\,-{\bf F}\,+\,
\partial {\bf U}/\partial t)/T \eqno(A2.8)$$
where ${\bf U}$ is the velocity of the gas particle,
$h_0=({\bf U \cdot U})/2+h$, ${\bf F}$ is the mass force.
The operator $grad$
in this equation is defined only in the plane normal to the trajectory.
[Here it was tolerated a certain incorrectness. Equations (A2.7), (A2.8) are 
written in different forms. This is connected with difficulties when
deriving these equations themselves.
However, this incorrectness will not effect on results of the qualitative
analysis of the evolutionary relation obtained from these equations.]

Since the total derivative with respect to time is that along the trajectory,
in the accompanying frame of reference equations (A2.7) and (A2.8)
take the form:
$${{\partial s}\over {\partial \xi ^1}}\,=\,0 \eqno (A2.9)$$
$${{\partial s}\over {\partial \xi ^{\nu}}}\,=\,A_{\nu },\quad \nu=2, ... \eqno(A2.10)$$
where $\xi ^1$ is the coordinate along the trajectory,
$\partial s/\partial \xi ^{\nu }$
is the left-hand side of equation (A2.8), and $A_{\nu }$ is obtained from the
right-hand side of relation (A2.8).

\{In the common case when gas is
nonideal equation (A2.9) can be written in the form
$${{\partial s}\over {\partial \xi ^1}} \,=\,A_1 \eqno (A2.11)$$
where $A_1$ is an expression that depends on the energetic actions 
(transport phenomena: viscous, heat-conductive). 
In the case of ideal gas $A_1\,=\,0$ and equation (A2.12) transforms 
into (A2.9). In the case of the viscous heat-conductive gas described by 
a set of the Navier-Stokes equations, in the inertial frame of reference 
the expression $A_1$ can be written as [13] 
$$A_1\,=\,{1\over {\rho }}{{\partial }\over {\partial x_i}}
\left (-{{q_i}\over T}\right )\,-\,{{q_i}\over {\rho T}}\,{{\partial T}\over {\partial x_i}}
\,+{{\tau _{ki}}\over {\rho }}\,{{\partial u_i}\over {\partial x_k}} \eqno(A2.12)$$
Here $q_i$ is the heat flux, $\tau _{ki}$ is the viscous stress tensor.
In the case of reacting gas extra terms connected with the chemical
nonequilibrium are added [13].\}

Equations (A2.9) and (A2.10) can be convoluted into the relation 
$$ds\,=\,A_{\mu} d\xi ^{\mu}\eqno(A2.13)$$
where $\,A_{\mu} d\xi ^{\mu}=\omega\,$ is the first degree differential 
form (here $A_1=0$,$\mu =1,\,\nu $).

Relation (A2.13) is the evolutionary relation for gas dynamic system
(in the case of local thermodynamic equilibrium). Here $\psi\,=\,s$.
$\{$It worth notice that in the evolutionary relation for thermodynamic
system the dependence of entropy on thermodynamic variables is investigated
(see relation (A2.5)), whereas in the evolutionary relation for gas dynamic
system the entropy dependence on the space-time variables is considered$\}$.

Relation (A2.13) appears to be nonidentical. To make it sure that this is true 
one must inspect the commutator of the form $\omega $.

Without accounting for terms that
are connected with a deformation of the manifold formed by the trajectories
the commutator can be written as
$$K_{1\nu }\,=\,{{\partial A_{\nu }}\over {\partial \xi ^1}}\,-\,{{\partial A_1}\over
{\partial \xi ^{\nu }}}$$

From the analysis of the expression $A_{\nu }$ and with taking into account 
that $A_1\,=\,0$ one can see that terms
that are related to the multiple connectedness of the flow domain (the second
term of  equation (A2.8)), the nonpotentiality of the external forces
(the third term in (A2.8)) and the nonstationarity of the flow (the forth term
in (A2.8)) contribute into the commutator. \{The terms 
connected with transport phenomena and physical and chemical processes will 
contribute into the commutator (see expression (A2.12)).\}

Since the commutator of the form $\omega $ is nonzero, it is evident that 
the form $\omega$ proves to be unclosed. 
This means that relation (A2.13) cannot be an identical one.

Nonidentity of the evolutionary relation points to the nonequilibrium 
state and the development of the gas dynamic instability.
Since the nonequilibrium state is produced by internal forces that are 
described by the commutator of the form $\omega $, it becomes evident 
that the cause of the gas dynamic instability is something that 
contributes into the commutator of the form $\omega $.

One can see (see (A2.8)) that the development of instability is caused by
not a simply connectedness of the flow domain,  nonpotential  external
(for each local domain of the gas dynamic system) forces, a nonstationarity
of the flow. 

All these factors lead to emergency of 
internal forces, that is, to nonequilibrium state and to development
of various types of instability. \{Transport 
phenomena and physical and chemical processes also lead to emergency of 
internal forces and to development of instability.\}

And yet for every
type of instability one can find the appropriate term giving contribution
to the evolutionary form commutator, which is responsible for this type
of instability.
Thus, there is an unambiguous connection between the type of instability
and the terms that contribute to  the evolutionary form commutator in the
evolutionary relation. \{In the general case one has to consider the
evolutionary relations that correspond to the balance conservation laws
for angular momentum and mass as well.\}

As it was shown above, under realization of additional degrees of freedom
it can take place the transition from the nonequilibrium state to the locally 
equilibrium one, and this process is accompanied by emergency of physical
structures.
The gas dynamic formations that correspond to these physical structures are
shocks, shock waves, turbulent pulsations and so on. Additional degrees of
freedom are realized as the condition of the degenerate transformation, namely,
vanishing of determinants, Jacobians of transformations, etc. These conditions
specify the integral surfaces (pseudostructures):
the characteristics (the determinant of coefficients at the normal derivatives
vanishes), the singular points (Jacobian is equal to zero), the envelopes
of characteristics of the Euler equations and so on. Under crossing
throughout the integral surfaces
the gas dynamic functions or their derivatives undergo the breaks.

Let as analyze which types of instability and what gas dynamic  formation
can originate under given external action.

1). {\it Shock, break of diaphragm and others}. The instability originates
because
of nonstationarity. The last term in equation (A2.8) gives a contribution
into the commutator. In the case of ideal gas whose flow is described by
equations of the hyperbolic type the transition to the locally equilibrium
state is possible on the characteristics and their envelopes. The
corresponding structures are weak shocks and shock waves.

2).{\it Flow of ideal (inviscous, heat nonconductive) gas around bodies
Action of nonpotential forces}. The instability develops because of
the multiple connectedness of the flow domain and a nonpotentiality of the
body forces. The contribution into the commutator comes from the second and
third terms of the right-hand side of  equation (A2.8). Since the gas is ideal
one and $\partial s/\partial \xi ^1=A_1=0$, that is, there is no contribution
into the each fluid particle, an instability of convective type develops.
For $U>a$ ($U$ is the velocity of the gas particle, $a$ is the speed of sound)
a set of equations of the balance conservation laws belongs to the
hyperbolic type and hence the transition to the locally equilibrium state is
possible on the characteristics and on the envelopes of characteristics
as well, and weak shocks and shock waves are the structures of the system.
If $U<a$ when the equations are of elliptic type, such a transition is
possible only at singular points. The structures emerged due to a convection
are of the vortex type. Under long acting the large-scale structures
can be produced.

3. {\it Boundary layer}. The instability originates due to the multiple
connectness of the domain and the transport phenomena (an effect of
viscosity and thermal conductivity). Contributions into the commutator produce
the second term in the right-hand side of equation (A2.8) and the second and
third terms in expression (A2.12). The transition to the locally equilibrium
state is allowed at singular points. because in this case
$\partial s/\partial \xi^1=A_1\neq 0$, that is, the external exposure acts
onto the gas particle separately, the development of instability and the
transitions to the locally equilibrium state are allowed only in
an individual fluid particle. Hence, the structures emerged behave as
pulsations. These are the turbulent pulsations.

\{Studying the instability on the basis of the analysis of entropy
behavior was carried out in the works by Prigogine and co-authors [23].
In that works entropy was considered as the thermodynamic function
of state (though its behavior along the trajectory was analyzed).
By means of such state function one can trace the development (in gas
fluxes) of the hydrodynamic instability only. To investigate the gas
dynamic instability it is necessary to consider entropy as the gas dynamic
state function, i.e. as a function of the space-time coordinates.
Whereas for studying the thermodynamic instability one has to analyze
the commutator constructed by the mixed derivatives of entropy with respect
to the thermodynamic variables, for studying the gas dynamic instability
it is necessary to analyze the commutators
constructed by the mixed derivatives of entropy with respect to the space-time
coordinates.\}

\subsection*{Electromagnetic field}

The system of charged particles is a material medium, which
generates  electromagnetic field.

If to use the Lorentz force ${\bf F\,= \,\rho (E + [U\times H]}/c)$,
the local variation of energy and linear momentum of the charged
matter (material system) can be written respectively as [10]: $\rho ({\bf U\cdot E})$,
$\rho ({\bf E+[U\times H]}/c)$. Here $\rho$ is the charge
density, ${\bf U}$ is the velocity of charged matter. These
variations of energy and linear momentum are caused by energetic and
force actions and are equal to values of these actions. If to denote
these actions by $Q^e$, ${\bf Q}^i$, the balance conservation laws
can be written as follows:
$$\rho \,({\bf U\cdot E})\,=\,Q^e\eqno(A2.14)$$
$$\rho \,({\bf E\,+\,[U\times H]}/c)\,=\, {\bf Q}^i \eqno(A2.15)$$

After eliminating  the characteristics of material system (the charged
matter) $\rho$ and ${\bf U}$ by using the Maxwell-Lorentz equations
[10], the left-hand sides of equations (A2.14), (A2.15) can be expressed only
in terms of the strengths of electromagnetic field, and then one can write
equations (A2.14), (A2.15) as
$$c\,\hbox{div} {\bf S}\,=\,-{{\partial}\over {\partial t}}\,I\,+\,Q^e\eqno(A2.16)$$
$${1\over c}\,{{\partial }\over {\partial t}}\,{\bf S}\,=
\,{\bf G}\,+\,{\bf Q^i}\eqno(A2.17)$$
where ${\bf S=[E\times H]}$ is the Pointing vector, $I=(E^2+H^2)/c$,
${\bf G}={\bf E}\,\hbox {div}{\bf E}+\hbox{grad}({\bf E\cdot E})-
({\bf E}\cdot \hbox {grad}){\bf E}+\hbox {grad}({\bf H\cdot H})-({\bf H}\cdot\hbox{grad}){\bf H}$.

Equation (A2.16) is widely used while describing electromagnetic
field and calculating  energy and the Pointing vector. But equation (A2.17)
does not commonly be taken into account. Actually, the Pointing vector
${\bf S}$ must obey two equations that can be convoluted into the
{\it relation}
$$d\bf S=\,\omega ^2\eqno(A2.18)$$
Here $d\bf S$ is the state differential being 2-form and the coefficients 
of the form $\omega ^2$ (the second degree form) are the right-hand sides 
of equations (A2.16) and (A2.17).
It is just the evolutionary relation for the system of charged particles that 
generate electromagnetic field. 

By analyzing the coefficients of the form $\omega ^2$ (obtained from equations 
(A2.16) and (A2.17), one can assure oneself that the form commutator is nonzero. 
This means that from relation (A2.18) the Pointing vector cannot be found.
This points to the fact that there is no such a measurable quantity
(a potential).

Under what conditions
can the Pointing vector be formed as a measurable quantity?

Let us choose the local coordinates $l_k$ in such a way that one direction 
$l_1$ coincides with the direction of the vector ${\bf S}$. Because this 
chosen direction coincides with the direction of the vector
${\bf S=[E\times H]}$ and hence is normal to the vectors
${\bf E}$ and ${\bf H}$,
one obtains that $\hbox{div} {\bf S}\,=\,\partial s/\partial l_1$,
where $S$ is a module of ${\bf S}$. In addition,
the projection of the vector ${\bf G}$ on the chosen direction turns out to be 
equal to $-\partial I/\partial l_1$.
As a result, after separating from vector equation (A2.17) its projection 
on the chosen direction equations (A2.16) and (A2.17) can be written as 
$${{\partial S}\over {\partial l_1}}\,=\,-{1\over c}{{\partial I}\over {\partial t}}\,+\,
{1\over c}Q^e \eqno(A2.19)$$
$${{\partial S}\over {\partial t}}\,=\,-c\,{{\partial I}\over {\partial l_1}}\,+\,c{\bf Q}'^i\eqno(A2.20)$$
$$0\,=\,-{\bf G}''\,-\,c{\bf Q}''^i$$
Here the prime relates to the direction $l_1$, double primes relate to 
the other directions. Under the condition $d l_1/d t\,=\,c$ from
equations (A2.19) and (A2.20) it is possible to obtain the relation in differential 
forms
$${{\partial S}\over {\partial l_1}}\,dl_1\,+\,{{\partial S}\over {\partial t}}\,dt\,=\,
-\left( {{\partial I}\over {\partial l_1}}\,dl_1\,+\,{{\partial I}\over {\partial t}}\,dt\right )\,+\,
(Q^e\,dt\,+\,{\bf Q}'^i\,dl_1)\eqno(A2.21)$$
Because the expression within the second braces in the right-hand side is 
not a differential (the energetic and force
actions have different nature and cannot be conjugated), one can obtain 
a closed form only if this term vanishes:
$$(Q^e\,dt\,+\,{\bf Q}'^i\,dl_1)\,=\,0\eqno(A2.22)$$
that is possible only discretely (rather than identically).

In this case $dS\,=\,0$, $dI\,=\,0$ and the modulus of the Pointing vector $S$ 
proves to be a closed form, i.e. a measurable quantity. The integrating 
direction (the pseudostructure) will be
$$-\,{{\partial S/\partial t}\over {\partial S/\partial l_1}}\,=\,{{dl_1}\over {dt}}\,=\,c\eqno(A2.23)$$
The quantity $I$ is the second dual invariant.

Thus, the constant $c$ entered into the Maxwell equations is defined
as the integrating direction.

1. Bott R., Tu L.~W., Differential Forms in Algebraic Topology. 
Springer, NY, 1982.

2. Encyclopedia of Mathematics. -Moscow, Sov.~Encyc., 1979 (in Russian).

3. Encyclopedic dictionary of the physical sciences. -Moscow, Sov.~Encyc., 
1984 (in Russian).

4. Dirac P.~A.~M., The Principles of Quantum Mechanics. Clarendon Press, 
Oxford, UK, 1958.

5. Wheeler J.~A., Neutrino, Gravitation and Geometry. Bologna, 1960.

6. Tonnelat M.-A., Les principles de la theorie electromagnetique 
et la relativite. Masson, Paris, 1959.

7. Pauli W. Theory of Relativity. Pergamon Press, 1958.

8. Schutz B.~F., Geometrical Methods of Mathematical Physics. Cambrige 
University Press, Cambrige, 1982.

9. Petrova L.~I. Properties of conservation laws and a mechanism 
of origination of physical structures. (The method 
of skew-symmetric differential forms). Ed. MSU, Moscow, 2002.  

10. Tolman R.~C., Relativity, Thermodynamics, and Cosmology. Clarendon Press, 
Oxford,  UK, 1969.

11. Fock V.~A., Theory of space, time, and gravitation. -Moscow, 
Tech.~Theor.~Lit., 1955 (in Russian).

12. Dafermos C.~M. In "Nonlinear waves". Cornell University Press,
Ithaca-London, 1974. 
      
13. Clark J.~F., Machesney ~M., The Dynamics of Real Gases. Butterworths, 
London, 1964. 

14. Petrova L.~I. Origination of physical structures. Izv. RAN, 
Ser.fizicheskaya, V.67, N3, pages 415-424.

15. Dirac P.~A.~M., Proc.~Roy.~Soc., {\bf A133}, 60 (1931).

16. Dirac P.~A.~M., Phys.~Rev., {\bf 74}, 817 (1948).

17. Petrova L.~I. Formation of physical fields and manifolds, 
//Proceedings International Scientific Meeting PIRT-2003 "Physical 
Interpretations of relativity Theory": Moscow, Liverpool, Sunderland, 
2004, pages 161-167. 

18. Einstein A. The Meaning of Relativity. Princeton, 1953.

19. Cartan E., Les Systemes Differentials Exterieus ef Leurs Application 
Geometriques. -Paris, Hermann, 1945.  

20. Smirnov V.~I., A course of higher mathematics. -Moscow, 
Tech.~Theor.~Lit. 1957, V.~4 (in Russian).

21. Haywood R.~W., Equilibrium Thermodynamics. Wiley Inc. 1980.

22. Liepman H.~W., Roshko ~A., Elements of Gas Dynamics. Jonn Wiley, 
New York, 1957

23. Glansdorff P., Prigogine I. Thermodynamic Theory of Structure, Stability 
and Fluctuations. Wiley, N.Y., 1971.  

\end{document}